\documentstyle[12pt]{article}
\newcommand{\be}{\begin{equation}}
\newcommand{\nn}{\nonumber}
\newcommand{\bea}{\begin{eqnarray}}
\newcommand{\eea}{\end{eqnarray}}
\newcommand{\ba}{\begin{array}}
\newcommand{\ea}{\end{array}}
\newcommand{\ee}{\end{equation}}

\expandafter\ifx\csname mathbbm\endcsname\relax

\else

\fi \textheight 22cm \textwidth 15cm \topmargin 1mm \oddsidemargin
5mm \evensidemargin 5mm

\newcommand{\beas}{\begin{eqnarray*}}
\newcommand{\eeas}{\end{eqnarray*}}
\newcommand{\bes}{\begin{equation*}}
\newcommand{\ees}{\end{equation*}}

\newcommand{\lf}{\left}
\newcommand{\ri}{\right}
\newcommand{\f}{\frac}
\newcommand{\dagg}{\dagger}

\def\tr           {\mbox{\rm tr}\,}

\def\i2           {\mbox{$\frac{i}{2}$}}

\def\al           {\alpha}

\def\bet           {\beta}
\def\beb           {{\bar \beta}}

\def\ft            {{\tilde F}_4}

\def\del           {\delta}

\def\ep           {\epsilon}

\def\et           {\eta}

\def\gab           {{\bar \gamma}}

\def\Om           {\Omega}
\def\ph           {\phi}

\def\ps           {\psi}

\def\rh           {\rho}

\def\si           {\sigma}

\def\th{\theta}

\def\pl           {\partial}

\def\we {{\wedge}}

\begin{document}

\begin{titlepage}
\vspace*{20mm}
\begin{center}
{\LARGE \bf{{$U(1)$ Instantons on $AdS_4$ and the Uplift to Exact Supergravity Solutions}}}\\ 

\vspace*{15mm} \vspace*{1mm} {Ali Imaanpur}

\vspace*{1cm}

{\it Department of Physics, School of Sciences\\ 
Tarbiat Modares University, P.O.Box 14155-4838, Tehran, Iran\\
Email: aimaanpu@theory.ipm.ac.ir}\\
\vspace*{1mm}

\vspace*{1cm}

\end{center}

\begin{abstract}
We consider self-duality equation of $U(1)$ gauge fields on Euclidean $AdS_4$ space, and find a simple 
finite action solution. With a suitable ansatz, we are able to embed this solution into the 10d supergravity 
background of $AdS_4\times CP^3$. Further, we show that the solution can be uplifted to an exact solution in 
11d supergravity background of $AdS_4\times SE_7$. In the context of ABJM model, we also discuss the coupling 
to the boundary conserved currents.  

\end{abstract}

\end{titlepage}

\section{Introduction}
Instantons have played an important role in understanding the nonperturbative effects in quantum field theories 
and string theory. AdS/CFT correspondence \cite{MAL}, on the other hand, has provided a new perspective on 
instantons in terms of D-brane solutions in string theory. In particular, Yang-Mills instantons have been identified with D(-1)-brane solutions in type IIB supergravity. Following this identification, it has been possible to trace 
over the corresponding nonperturbative effects on both sides of the duality, and hence testing the AdS/CFT 
duality beyond the perturbative level \cite{BIA, BAN}. 

The AdS/CFT correspondence has further been generalized by Aharony, \linebreak 
Bergman, Jafferis, and Maldacena (ABJM) to M-theory (and type IIA theory upon compactification) \cite{ABJM}. 
Therefore, to investigate the nonperturbative characteristics of this duality, it is important to look for 
some exact D-brane solutions in the corresponding supergravity backgrounds, and try to identify the dual 
instantons on the boundary Chern-Simons theory. Recently, we succeeded in constructing such dual instanton 
configurations in the antimembranes theory ignoring the backreaction on the metric. This construction 
further led us to propose that the antimembranes boundary theory is related to the ABJM model by swapping 
the $\bf{s}$ and $\bf{c}$ representations of the $SO(8)$ global symmetry \cite{ALI}. In the present paper, 
however, we provide the first examples of {\em exact} solutions on $AdS_4\times CP^3$, and $AdS_4\times SE_7$ 
for the type IIA and M-theory backgrounds, respectively. 

We start with the self-duality equation on $AdS_4$ space, and find a solution which has a finite action. 
Because of the self-duality, the energy-momentum tensor of this solution vanishes and hence there will be 
no backreaction on the metric. This implies that we have in fact an exact solution to the equations 
of motion coming from the Maxwell-Einstein action. In Sec. 3, we provide an ansatz for a system of 
D0-D2 brane configuration and discuss how the self-dual gauge fields can be embedded into an exact solution 
of 10d supergravity on $AdS_4\times CP^3$. As in four dimensions, the energy-momentum tensor of individual 
branes along $AdS_4$ vanishes, whereas the components along $CP^3$ add up to zero. Hence the background metric 
will not change in the presence of branes (fluxes).  In Sec. 4, we use the consistent truncation of \cite{GAUNT} 
to uplift our 4-dimensional solution to an exact solution on $AdS_4\times SE_7$. We will see explicitly how 
this comes about by looking at the reduced four-dimensional equations. In Sec. 5, we discuss the coupling to the 
boundary operators. We will determine the dual operators by examining the symmetry properties of the supergravity ansatzs under the isometry group of the metric.

\section{$U(1)$ Instantons on $AdS_4$}
To discuss the self-dual gauge fields on Euclidean $AdS_4$,\footnote{$U(1)$ instantons on $AdS_4$ have also been 
discussed in \cite{DEH}. Here, however, we take a different approach.} 
we use the Poincare coordinates for the metric:
\be
ds^2= \f{1}{\rh^2}(d\rh^2 + dx_1^2 + dx_2^2 + dx_3^2) \, ,
\ee
which reflects the conformal flatness of the metric. On the other hand, the (anti)self-duality 
condition is invariant under the conformal transformations of the metric, so on $AdS_4$ we can write 
\be
F_{\mu\nu}=-\f{\sqrt{g}}{2}\,\ep_{\mu\nu\rh\si}F^{\rh\si}=
-\f{1}{2}\,\ep_{\mu\nu\rho\si}\del^{\rho\xi}\del^{\si\eta}F_{\xi\eta}\, ,\label{SELFD}
\ee
just as on flat $\bf{R}^4$.  

To find a solution, we make the following ansatz for the $U(1)$ gauge field $A_\mu$:
\be
A_1=x_2 h(r,\rh)\, , \ \ \ \ A_2=-x_1 h(r,\rh)\, ,\ \ \ \ A_3=g(r,\rh)\, , \ \ \ 
A_4=-x_3 h(r,\rh)\, , \label{ANS}
\ee
where $r=\sqrt{x_1^2 +x_2^2 + x_3^2}$. Note that this ansatz respects the $SO(3)$ symmetry 
along the 3-dimensional space orthogonal to the radial direction. It also resembles an ansatz 
used in \cite{SEI} for solving the noncommutative $U(1)$ instanton equation on ${\bf R}^4$. 
For the field strengths we get
\bea
&& F_{12}=-2h -\f{h'}{r}(x_1^2+x_2^2)\, ,\ \ \ F_{34}=-h-\f{h'}{r}x_3^2 -\dot{g}\nn \\
&& F_{23}=\f{1}{r}(g' x_2  +h' x_1x_3 )\, ,\ \ \ F_{14}=\f{-h'}{r}x_1x_3 -x_2\dot{h}\nn \\
&& F_{24}=-\f{h'}{r}x_2x_3+x_1 \dot{h}\, ,\ \ \ F_{31}=\f{h'}{r}x_2x_3 -\f{g'}{r}x_1\, ,\nn 
\eea  
where prime and dot indicate the differentiation with respect to $r$ and $\rho$, respectively. 
Now let us impose the self-duality conditions. From $F_{12}=-F_{34}$ we obtain
\be
-3h-rh'=\dot{g}\, ,\label{g}
\ee
or
\be
\f{1}{r^2}\f{\pl}{\pl r}(r^3 h)+\f{\pl g}{\pl \rh}=0\, , \label{h}
\ee
while $F_{23}=-F_{14}$, or $F_{24}=-F_{31}$ yield
\be
\dot{h}=\f{g'}{r}\, .\label{g1}
\ee 
Taking the derivative of the above equation with respect to $\rh$ and using (\ref{h}), we have
\be
\f{\pl }{r\pl r}\lf[\f{1}{r^2}\f{\pl}{\pl r}(r^3 h)\ri] +\f{\pl^2 h}{\pl \rh^2}=0\, .\label{h2}
\ee

Let us further make an assumption that $h$ depends on $r$ and $\rh$ through $z\equiv (\rh+\rh_0)^2 + r^2$, i.e.,
\be
h=h(z)=h((\rh+\rh_0)^2 + r^2)\, .
\ee
Plugging this ansatz into (\ref{h2}), we get
\be
z\, \f{\pl^2 h}{\pl z^2} + 3\, \f{\pl h}{\pl z}=0\, ,
\ee
which has the simple solution of
\be
h(r,\rh)= C_1 +\f{C_2}{[(\rh+\rh_0)^2+r^2]^2}\, .\label{SOL}
\ee
Note that with $\rh_0 \neq 0$ we will have a smooth solution. Actually here lies the difference with 
the $U(1)$ instantons on flat $\bf{R}^4$, where there is no way to avoid singularities of the solution. 
Furthermore, as we will see presently, with $\rh_0$ in (\ref{SOL}) the action turns out to be finite. 
Let us further set $C_1=0$, then we can write
\be
g(r,\rh)=(\rh +\rh_0)\, h(r,\rh)\, ,
\ee
which satisfies both (\ref{g}) and (\ref{g1}). The full moduli of the solution can be seen by writing  
$A_\mu$ as
\bea
&& A_1=(x_2-x_2^0) h(r,\rh)\, , \ \ \ \ A_2=-(x_1-x_1^0) h(r,\rh)\, ,\nn \\
&& A_3=(\rh +\rh_0)\, h(r,\rh)\, , \ \ \ A_4=-(x_3-x_3^0) h(r,\rh)\, ,\label{SELF}
\eea 
now with $r^2=(x_1-x_1^0)^2+(x_2-x_2^0)^2+(x_3-x_3^0)^2$, and hence
\be
h(r,\rh)= \f{C_2}{[(\rh+\rh_0)^2+(x_1-x_1^0)^2+(x_2-x_2^0)^2+(x_3-x_3^0)^2]^2}\, .
\ee
So we have (\ref{SELF}) as our solution of the self-duality equation on $AdS_4$.

\subsection{The Boundary Term}
To complete our discussion, let us look at the variation of the action in the presence of a boundary. This proves 
to be useful in our study of the coupling to the boundary operators of the dual CFT. For the variation of the  
Maxwell action we have
\bea
\del S &=&\f{1}{2} \int \, d^4x\, F_{\mu\nu}\, \del F^{\mu\nu} \nn \\
&=& -\int \, d^4x\, \del A_\nu (\pl_\mu F^{\mu\nu}) + \int \, d^4x\, \pl_\mu( \del A_\nu\, F^{\mu\nu})\nn \\
&=& -\int \, d^4x\, \del A_\nu (\pl_\mu F^{\mu\nu}) - \, (\del A_i\, F^{\rho i})|_{\rho =0}\, ,
\eea
here $i,j,\ldots =1,2,3,$ indicate the boundary tangent indices, and $\rho$ is the radial direction so that 
the boundary is at $\rho =0$. So for having the equation of motion in the bulk, on the boundary we have to 
have either $\del A_i=0$ (Dirichlet boundary condition) or $F_{\rho i}=0$ (Neumann boundary condition). 
However, neither of these boundary conditions are consistent with self-duality condition (\ref{SELFD}) 
in the bulk. For this to happen, we deform the Maxwell action as follows:
\be
S =\f{1}{4}\int \, d^4x\, F_{\mu\nu}F^{\mu\nu} -\f{1}{4}\int d^3x\, \ep_{ijk}\, A^i F^{jk}\, ,
\ee
so that the variation of the action now reads
\be
\del S = -\int \, d^4x\, \del A_\nu (\pl_\mu F^{\mu\nu}) - \, \del A_i\, 
(F^{\rho i}+\f{1}{2}\ep_{ijk}F^{jk})|_{\rho =0}\, .
\ee
Now, on the boundary we can demand
\be
F^{\rho i}=-\f{1}{2}\, \ep_{ijk}\, F^{jk}\label{MIX}
\ee 
which is also consistent with the self-duality condition in the bulk. This is a sort of mixed boundary 
condition as it relates the electric field to the gauge invariant part of the gauge potential. 

\subsection{The Action}
With our ansatz (\ref{ANS}), we can compute the action of instantons on $AdS_4$:
\bea
S&=&\f{1}{4}\int \sqrt{g}\, d^4x\, F_{\mu\nu}F^{\mu\nu}=-\f{1}{8}\int \, d^4x\, \ep_{\mu\nu\rh\si}F^{\mu\nu}F^{\rh\si}\nn \\
&=&-\f{1}{2}\int  d^4x\, (F_{12}F_{34}+F_{23}F_{14}+F_{24}F_{31}) \nn \\
&=& \f{1}{2}\int d^4x \lf[4h^2 + (x_1^2+x_2^2)(h'^2+\dot{h}^2 +\f{4hh'}{r})\ri] \nn \\
&=& 4\pi\int r^2 dr d\rh \lf[2h^2 + \f{r^2}{3}(h'^2+\dot{h}^2 +\f{4hh'}{r})\ri]\, , \nn 
\eea
where in the last line we have performed the integral over $\th$ and $\phi$. Using solution (\ref{SOL}),  
with $C_1=0$, results in
\bea
S= 8\pi C_2^2 \int \f{r^2dr d\rh}{[(\rh+\rh_0)^2+r^2]^4}
= \f{\pi^2 C_2^2}{2}  \int_0^\infty \f{d\rh}{2(\rh+\rh_0)^5}
= \f{\pi^2 C_2^2}{16\rh_0^4}\, .\label{ACT}
\eea
If $\rh_0$ is going to be a modulus of the solution, we need to choose $C_2\sim \rh_0^2$. This makes $S$ 
independent of $\rh_0$,  and gives $A_\mu$ the right dimension of one. We have therefore obtained a finite action solution 
of equations of motion, i.e., a $U(1)$ instanton in $AdS_4$. Moreover, being self-dual, the solution 
has a vanishing energy-momentum tensor and hence it provides an exact solution to the Maxwell-Einstein 
equations in four dimensions.

\section{Uplift to 10d Type IIA Supergravity Solutions}
In this section we will see how the self-dual gauge fields can be embedded into a solution of type IIA supergravity. 
In fact, our construction of $U(1)$ instantons on $AdS_4$ in the previous section was motivated by our search for an 
exact type IIA solution on $AdS_4\times CP^3$ background. 

To begin the discussion, let us recall the Euclidean action of type IIA 
supergravity in the string frame
\bea
S_{IIA}&=&\f{1}{2\kappa^2}\int d^{10}x\, e^{-2\ph}\sqrt{ g}\, R 
+\f{1}{2\kappa^2}\int \lf(e^{-2\ph}(4d\ph\wedge *d\ph -\f{1}{2}H\wedge *H)\ri. \nn \\ 
&&\lf. -\f{1}{2}F_2\wedge *F_2
-\f{1}{2}\ft\wedge *\ft +\f{i}{2}B\wedge F_4\wedge F_4 \ri) \, .
\eea
For the field equations we have
\bea
&& d\ft=-F_2\wedge H \, , \ \ \ \ d*\ft=-\ft\wedge H \, , \ \ \ \ dH=0\, , \ \ \ \ dF_2=0\, ,\nn \\ 
&& d*(e^{-2\ph}H)=-F_2\wedge *\ft-\f{i}{2}\ft\wedge \ft \, , \ \ \ \ d*F_2=H\wedge *\ft \, , \nn \\
&& d*d\ph-d\ph\wedge *d\ph -\f{1}{8}H\wedge *H +\f{1}{4\cdot 3!}R\, \ep_4\we J^3=0\, ,\nn \\
&& \ft=F_4-A_1\wedge H \, .\label{FEQ}
\eea
With the Euclidean signature, we have the following background solution\footnote{We have explicitly 
checked for the factor of 2 in $F_2$.}
\bea
&& ds^2=\f{R^3}{k}(\f{1}{4}ds_{AdS_4}^2 +ds^2_{CP^3}) \, , \nn \\
&& e^{2\phi}=\f{R^3}{k^3}\, ,\ \ \ F_4=-\f{3i}{8}R^3\ep_4\, ,\ \ \ \ F_2=2kJ\, ,\label{BAC}
\eea 
where $J$ is the K\"ahler form on $CP^3$. 

Having had the background solution (\ref{BAC}), we would like to look for a new solution on this background, 
i.e., a D-instanton.  So to proceed, let us make the following ansatz:
\be
F_2 =2kJ+F \, , \ \ \ \ F_4 =-\f{3i}{8}R^3\ep_4 +i\al\, J\we F\, ,\ \ \ \ H=0 \, ,\label{AN}
\ee
with $\al$ a constant parameter, and $F$ a 2-form in $AdS_4$. The extra terms in $F_2$ and $F_4$ can be 
thought to be sourced by a D0-brane and D2-brane, respectively. 

Setting $H=0$, the field equations (\ref{FEQ}) now read
\bea
&&dF_4 =0\, ,\ \ \ \ d*F_4=0\, ,\ \ \ \ dF_2=0\, , \ \ \ \ d*F_2=0\, , \nn \\
&& 0=d*(e^{-2\ph}H)=-F_2\wedge *F_4 -\f{i}{2}F_4\wedge F_4\, . \label{SUG}
\eea
As we will discuss in the following, the extra terms in our ansatz (\ref{AN}) will have no backreaction on 
the metric so that the background dilaton $\ph$ continues to satisfy the field equations.  
If we choose $F$ to be a self-dual 2-form in $AdS_4$, the field equations of $F_2$ and $F_4$ are obviously 
satisfied. What about the last equation in (\ref{SUG})?  Let us expand the right hand side of this 
equation:\footnote{Note that $ \ep_4=vol(AdS_4)\, , \ vol(CP^3)=\f{1}{3!}J\wedge J\wedge J\equiv 
\f{1}{3!}J^3\, ,\ \  *_6J = \f{1}{2}J\wedge J\equiv \f{1}{2}J^2$, and  $J\wedge J^3=0 \, .$ We use $*_4$ 
and $*_6$ to indicate the Hodge star operation with respect to the metrics of $AdS_4$ and $CP^3$, respectively.}
\bea
&&F_2\wedge *F_4 +\f{i}{2}F_4\wedge F_4 =\nn \\
&&=(2kJ+ F)\wedge *\lf(-\f{3i}{8}R^3\ep_4 +i\al\, J\we F\ri)
-\f{i\al^2}{2}\, F\wedge F\we J^2\nn \\ 
&&= (2kJ+ F)\wedge \lf(-\f{iR^6}{k}\, J^3 + \f{i\al}{2}\, J^2\wedge *_4 F \ri) -\f{i\al^2}{2}\, F\wedge F\we J^2\nn \\ 
&&= i\al R^3 *_4F\we J^3 -\f{iR^6}{k} F\we J^3 +\f{i\al R^3}{2k}\, F\we *_4 F\we J^2
-\f{i\al^2}{2}\, F\wedge F\we J^2. \label{SEQ}
\eea
Hence, for (\ref{SEQ}) to vanish we must have
\be
F=*_4 F\, , 
\ee
and
\be
\al =\f{R^3}{k}\, .
\ee 

The nice thing about ansatz (\ref{AN}) is that the indices of $F$ and $J\we F$ do not contract with 
those of the background fields. So in discussing the energy-momentum tensors, we need only to be concerned 
with the contributions of the these terms. Further, with $F$ a self-dual 
2-form the energy-momentum tensor along $AdS_4$ vanishes. Let $\mu , \nu ,\ldots$ and $\al ,\bet ,\ldots$ 
indicate the tangent indices on $AdS_4$ and $CP^3$, respectively, then we have
\be
T_{\mu\nu}^{F_2}=\f{1}{2\cdot 2!}\lf[2F_{\mu\rh}F_\nu^{\ \rh}-\f{1}{2}g_{\mu\nu}F_{\et\rh}F^{\et\rh}\ri]=0 
\ee
as $F$ is self-dual. For $F_4$ we have 
\bea
T_{\mu\nu}^{F_4}&=&-\f{1}{2\cdot 4!}\f{\al^2 k^2}{R^6}\lf[4\cdot 3 F_{\mu\rh\al\bet}F_\nu^{\ \rh\al\bet}-
\f{1}{2}\cdot 6\,  g_{\mu\nu}F_{\et\rh\al\bet}F^{\et\rh\al\bet}\ri] \nn \\
&=&-\f{3}{2\cdot 4!}\lf[4F_{\mu\rh}F_\nu^{\ \rh}-
 g_{\mu\nu}F_{\et\rh}F^{\et\rh}\ri] 2J_{\al\beb}J^{\al\beb}\nn \\
&=&-\f{3}{4}\lf[2F_{\mu\rh}F_\nu^{\ \rh}-\f{1}{2}g_{\mu\nu}F_{\et\rh}F^{\et\rh}\ri]=0 \, .
\eea

If we use the complex coordinate on $CP^3$, for the energy-momentum tensor of $F_2$ along $CP^3$ 
we have
\be
T_{\al\beb}^{F_2}=-\f{1}{8}\f{R^3}{k}g_{\al\beb}F_{\mu\nu}F^{\mu\nu} 
\ee
whereas, for $F_4$ it reads
\bea
T_{\al\beb}^{F_4}&=&-\f{1}{2\cdot 4!}\f{\al^2 k}{R^3}\lf[4\cdot 3 F_{\mu\rh}F^{\mu\rh}J_{\al\gab}
J_{\beb}^{\ \gab}-\f{1}{2}\cdot 6\, g_{\al\beb}F_{\et\rh}F^{\et\rh}J^2\ri] \nn \\
&=& +\f{1}{8}\f{R^3}{k}g_{\al\beb}F_{\mu\nu}F^{\mu\nu}\, ,
\eea
where we used
\be
J_{\al\gab}J_{\beb}^{\ \gab}=g_{\al\beb}\, ,\ \ \ \ \ J_{\al\beb}J^{\al\beb}=3 \, .
\ee
Therefore we conclude that
\be
T_{\al\beb}^{F_2}\ +\ T_{\al\beb}^{F_4}\ = 0 \, .
\ee 
So we have showed that with ansatz (\ref{AN}) the energy-momentum tensor of 
D0-D2 brane configuration vanishes and there will be no backreaction on the 
background metric, i.e., the Einstein equations are the same as before. Also note 
that the dilaton has its original value of $e^{2\ph} =R^3/k^3$. In conclusion, 
we have obtained an exact solution of the form (\ref{AN}) with $F$ a 
self-dual gauge field explicitly constructed in Sec. 2.

\section{Uplift to 11d Supergravity Solutions}
The $U(1)$ instanton that we found in Sec. 2, can further be uplifted to an exact solution in 
11d supergravity. Fortunately, the consistent truncation of 11d supergravity to four dimensions 
of \cite{GAUNT} makes this uplift much simpler, and in fact it is more general than our discussion in the 
previous section. To begin with, consider the following ansatz for the metric of 11d  
supergravity
\be
ds^2 = ds^2_4 + e^{2U} ds^2_{KE_6} + e^{2V}(\et + A_1)\otimes (\et + A_1) \, ,\label{MET}
\ee
where $ds^2_4$ is an arbitrary metric on a 4d spacetime, and $ds^2_{KE_6}$ is the metric 
of a 6d K\"ahler-Einstein space. $U$ and $V$ are scalar fields and $A_1$ is a one-form defined on the 
four-dimensional space. If we set $U=V=A_1=0$, then the the last two factors of the metric constitute 
the Sasaki-Einstein metric
\be
ds^2_{SE_7} = ds^2_{KE_6} + \et\otimes \et \, ,
\ee 
where $\et$ is the globally defined one-form dual to the Reeb Killing vector. 
On a Sasaki-Einstein manifold one can globally define a 2-form $J$ and a holomorphic 3-form 
so that
\be
d\et = 2J\, ,\ \ \ \ \ d\Om =4i\et\we\Om \, .
\ee
The 4-form $F_4$ is taken to be
\bea 
F_4 &=& f {\rm vol}_4 + H_3\we (\et + A_1) + H_2\we J + H_1 \we J \we (\et + A_1)\nn \\
&+& 2h\, J\we J + \sqrt{3}\, [\chi_1\we\Om + \chi (\et+A_1)\we \Om + c.c.] \, ,\label{G}
\eea
where $f$ and $h$ are real scalars, $H_p$, $p = 1, 2, 3$, are real $p$-forms, $\chi_1$ is a complex
one-form, and $\chi$ is a complex scalar on the four-dimensional spacetime. With the ansatzs  
(\ref{MET}) and (\ref{G}), the 11-dimensional supergravity field equations all reduce to 4-dimensional 
equations, so that any solution to the reduced equations can be lifted to an 11-dimensional supergravity 
solution. These 4-dimensional equations are given in \cite{GAUNT}. 

Now, as mentioned in that paper one can further consistently truncate by setting $U=V=h=\chi=H_3=H_1=0$. 
Equations (B.9) to (B.11) of \cite{GAUNT}, when adapted to Euclidean signature, reduce to 
\bea
&&-fF+6*_4H=0 \, ,\nn \\  
&&\ d*_4H=0\, , \nn \\  
&&*_4H\we F+iH\we H=0\, ,\label{H1}
\eea
with $F=dA_1$, $H\equiv H_2$, and $dH=0$. Equations (B.19), (B.20), and (B.22) reduce to
\bea
&&R_{\mu\nu}=\f{1}{3}g_{\mu\nu}f^2 +\f{1}{2}F_{\mu\rh}F_{\nu}^{\ \rh} +\f{3}{2}(H_{\mu\rh}H_\nu^{\ \rh}-\f{1}{6}g_{\mu\nu}H_{\rh\si}H^{\rh\si})
\, ,\nn \\
&&\ d*_4F=0\, , \nn \\
&&F_{\mu\nu}F^{\mu\nu}+H_{\mu\nu}H^{\mu\nu}=0\, .\label{H2} 
\eea
The last eqs. of (\ref{H1}) and (\ref{H2}) are satisfied if we set 
\be
H=i*_4F\, ,\label{STA}
\ee
from which $d*_4H=0$ and $d*_4F=0$ are followed by the Bianchi identities for $F$ and $H_2$, respectively. 
The first equation in (\ref{H1}) then implies $f=6i$. We are then left with the first equation in (\ref{H2}):
\bea
R_{\mu\nu}&=&-12 g_{\mu\nu} +\f{1}{2}F_{\mu\rh}F_{\nu}^{\ \rh} +\f{3}{2}(H_{\mu\rh}H_\nu^{\ \rh}-\f{1}{6}g_{\mu\nu}H_{\rh\si}H^{\rh\si})\nn \\
&=&-12 g_{\mu\nu} +2(F_{\mu\rh}F_\nu^{\ \rh}-\f{1}{4}g_{\mu\nu}F_{\rh\si}F^{\rh\si})\, ,\label{R}
\eea
where we have used
\be
H_{\mu\rh}H_\nu^{\ \rh}=F_{\mu\rh}F_\nu^{\ \rh}-\f{1}{2}g_{\mu\nu}F_{\rh\si}F^{\rh\si}\, ,
\ee
implied by (\ref{STA}). 

As a final step, let us set
\be
F=*_4F\, ,
\ee
for which the second term in (\ref{R}) (i.e., the energy-momentum tensor of $F$) vanishes, and hence 
equation (\ref{R}) is solved when the metric is that of Euclidean $AdS_4$.

\section{Coupling to the Boundary Operators}
The supergravity backgrounds we have considered so far appear in the ABJM model, where a concrete CFT 
dual has been proposed. In this model a Chern-Simons-matter theory describes the low energy dynamics of 
$N$ $M2$-branes at the tip of the orbifold ${\bf C}_4/{\bf Z}_k$. This theory is in turn conjectured to 
be dual to M-theory on $AdS_4\times S^7/{\bf Z}_k$, with $k$ the level of the Chern-Simons term in the 
gauge theory side. For large $k$ ($k^5 >>  N$), the dual theory is better described in terms of type IIA 
string theory on $AdS_4\times CP^3$ \cite{ABJM}. In this set up, we would now like to examine the effect 
on the boundary of turning on the gauge fields in the bulk. 

According to the standard prescription of AdS/CFT, the boundary value of a bulk mode acts as a source for 
the dual operator on the boundary theory \cite{WIT}. In particular, the boundary value of a gauge field  
couples to a dimension 2 conserved current $J_i$ through 
\be
\int d^3x\, A_i J^i\, .\label{AJ}
\ee
This is the case when we are imposing Dirichlet boundary conditions. 
For Neumann boundary conditions, on the other hand, the electric field is held fixed on the boundary 
and thus couples to gauge fields which are the boundary value of the dynamical bulk gauge field. 
In fact, Neumann boundary conditions are related to Dirichlet boundary conditions through 
the action of $SL(2,{\bf Z})$ duality in the bulk. The generator $S$ of this group acts through
\be
\lf(
\begin{array}{clrr}      
\vec{ E}    
 \\
\vec{B}
\end{array}\ri)\, 
\to  \lf(
\begin{array}{clrr}      
0 &  1      
 \\
-1  & 0
\end{array}\ri)\lf(
\begin{array}{clrr}      
\vec{E}      
 \\
\vec{B}
\end{array}\ri)= \lf(
\begin{array}{clrr}      
\vec{B}      
 \\
-\vec{E}
\end{array}\ri)\, ,
\ee
so in terms of the transformed gauge fields, Dirichlet boundary condition of $\vec{B}=0$ corresponds in terms of 
original gauge fields to Neumann boundary condition of $\vec{E}=0$ \cite{WSL}. The mixed boundary condition 
in (\ref{MIX}), however, can be realized by the application of $ST$. Let $T$ denote the second generator of 
$SL(2,{\bf Z})$
\be
T=\lf(
\begin{array}{clrr}      
1 &  1      
 \\
0 & 1
\end{array}\ri) 
\ee
so that $ST$ acts through
\be
\lf(
\begin{array}{clrr}      
\vec{E}      
 \\
\vec{B}
\end{array}\ri)\, 
\to   \lf(
\begin{array}{clrr}      
\vec{B}      
 \\
-\vec{E}-\vec{B}
\end{array}\ri)\, ,
\ee
we see that the Dirichlet condition ($\vec{B}=0$) of the transformed gauge fields corresponds in terms of the original 
gauge fields to the mixed boundary condition 
\be
\vec{E}+\vec{B}=0\, ,
\ee
which is the same boundary condition as (\ref{MIX}). The above $ST$ operation of $SL(2,{\bf Z})$ on the bulk 
gauge fields translates on the boundary to the following change of the Dirichlet coupling:
\be
\int d^3x\, A_iJ^i \to \int d^3x\, \lf(A_iJ^i + \f{1}{2\pi} \ep^{ijk}C_i \pl_j A_k  + \f{1}{4\pi} \ep^{ijk}A_i \pl_j A_k\ri)\, .\label{AJ2}
\ee
The last term is the effect of $T$ operation, whereas $S$ operation introduces a dual background gauge field 
$C_i$ and couples it to the conserved current $\ep^{ijk}\pl_j A_k$ and then promotes $A_i$ to a dynamical 
field \cite{WSL}.  

So far, we have determined boundary coupling (\ref{AJ2}) which is induced by the bulk self-dual gauge fields. 
The coupling, however, depends on the current $J$ which we are going to discuss next. 
In the ABJM model, a $U(N)\times U(N)$ Chern-Simons-matter theory describes the dynamics on the boundary. 
Let $A_i$ and $\hat{A}_i$ indicate the corresponding gauge fields of the $U(N)$ factors. The matter content 
consists of bosonic fields $Y^A$, and fermionic fields $\ps_A$ transforming in the ${\bf 4}$ and ${\bar {\bf 4}}$ 
representations of the $SU(4)$ global symmetry, respectively. The global symmetry is in fact $SU(4)\times U(1)_b$, which is enhanced to $SO(8)$ for $k=1,2$. 
Further, for the $U(1)$ parts, let $A^\pm_i = \tr (A_i\pm {\hat A}_i)$, the matter field then couples only 
to $A^-$, whereas $A^+$ appears in the action through
\be
\f{k}{4\pi}\int \ep^{ijk}A^-_{i}F^+_{jk}\, .
\ee
Hence, apart from $U(1)_b$ symmetry, which is generated by
\be
J_i= i\, \tr \lf(Y^\dagg _A D_i Y^A -  Y^A D_i Y^\dagg _A\ri)
\ee
there is a further global symmetry associated to the shift symmetry of $A^+$ which is generated by the current 
$\tilde{J}=*F^+$. So, altogether, we can identify two global $U(1)$ symmetries in the theory. 

Let us then see to which bulk gauge excitations these currents could couple. A look back at ansatzs (\ref{MET}) 
and (\ref{G}) shows that they are only invariant under $SU(4)\times U(1)$ subgroup of $SO(8)$ isometry 
group (the global symmetry when $k=1,2$). So, the excitations must couple to $J_i$, which is also invariant under 
$SU(4)\times U(1)$ but not the full $SO(8)$ symmetry group. Notice that ${\tilde J}_i$ is invariant under $SO(8)$,  
so it cannot couple to the bulk gauge field excitations we considered in this paper.

\pagebreak

\hspace{30mm}


\vspace{1.5mm}

\noindent

\vspace{1.5mm}

\noindent

\newpage



\end{document}